\documentstyle[aps,preprint,epsf]{revtex}

\begin{document}
\draft
\title
{\bf Superconductivity in the Pseudogap State in ``Hot -- Spots'' model:\ 
Ginzburg -- Landau Expansion}
\author{E.Z.Kuchinskii,\ M.V.Sadovskii,\ N.A.Strigina}
\address
{Institute for Electrophysics,\ \\ 
Russian Academy of Sciences,\ Ural Branch,\ \\ Ekaterinburg,\ 620016, Russia\\
E-mail:\ kuchinsk@iep.uran.ru,\ sadovski@iep.uran.ru,\ strigina@iep.uran.ru} 
\maketitle


\begin{abstract}
We analyze properties of superconducting state (for both $s$-wave and
$d$-wave pairing), appearing on the ``background'' of the pseudogap state,
induced by fluctuations of ``dielectric'' (AFM(SDW) or CDW) short -- range
order in the model of the Fermi surface with ``hot spots''.
We present microscopic derivation of Ginzburg -- Landau expansion, taking
into account all Feynman diagrams of perturbation theory over electron
interaction with this short -- range order fluctuations, leading to strong
electronic scattering in the vicinity of ``hot spots''. We determine the
dependence of superconducting critical temperature on the effective width
of the pseudogap and on correlation length of short -- range order 
fluctuations. We also find similar dependences of the main characteristics
of such superconductor close to transition temperature. It is shown 
particularly, that specific heat discontinuity at the transition temperature
is significantly decreased in the pseudogap region of the phase diagram.
\end{abstract} 

\newpage
\narrowtext

\section{Introduction}

Pseudogap state observed in a wide region of the phase diagram of high --
temperature superconducting cuprates is characterized by numerous anomalies
of their properties, both in normal and superconducting states \cite{Tim,MS}. 
Apparently, the most probable scenario of pseudogap state formation in
HTSC -- oxides can be based \cite{MS} on the picture of strong scattering of 
current carriers by fluctuations of short -- range order of ``dielectric'' type
(e.g. antiferromagnetic (AFM(SDW)) or charge density wave (CDW)) existing in
this region of the phase diagram. In momentum space this scattering takes
place in the vicinity of characteristic scattering vector
${\bf Q}=(\frac{\pi}{a},\frac{\pi}{a})$ ($a$ -- lattice constant),
corresponding to doubling of lattice period (e.g. vector of antiferromagnetism),
being a ``precursor'' of spectrum transformation, appearing after the
establishment of AFM(SDW) long -- range order. Correspondingly, there appears
an essentially non Fermi -- liquid like renormalization of electronic spectrum
in the vicinity of the so called ``hot spots'' on the Fermi surface \cite{MS}.
Recently there appeared a number of experiments giving rather convincing
evidence for precisely this scenario of pseudogap formation \cite{Lor,Kras,Arm}.
Within this picture it is possible to formulate simplified ``nearly exactly''
solvable model of the pseudogap state, describing the main properties of this
state \cite{MS}, and taking into account all Feynman diagrams of perturbation
theory for scattering by (Gaussian) fluctuations of (pseudogap) short -- range
order with characteristic scattering vectors from the area of ${\bf Q}$, 
with the width of this area defined by the appropriate correlation length $\xi$ 
\cite{Sch,KS}.

Up to now the majority of theoretical papers is devoted to the studies of 
models of the pseudogap state in the normal phase for $T>T_c$. In Refs.
\cite{SP,KS1,KS2,KS3} we have analyzed superconductivity in the simplified
model of the pseudogap state, based on the assumption of existence of ``hot''
(flat) patches on the Fermi surface. Within this model we have constructed
Ginzburg -- Landau expansion for different types (symmetries) of Cooper
pairing \cite{SP,KS2} and also studied the main properties of superconducting
state for $T<T_c$, solving the appropriate Gor'kov's equations 
\cite{KS1,KS2,KS3}. At first stage, we have considered greatly simplified 
(``toy'') model of Gaussian fluctuations of short -- range order with an 
infinite correlation length, where it is possible to obtain an exact (analytic)
solution for the pseudogap state \cite{SP,KS1}, further analysis of more
realistic case of finite correlation lengths was performed both for
``nearly exactly'' solvable model of Ref. \cite{KS2} (assuming the self --
averaging nature of superconducting order parameter over pseudogap 
fluctuations) and for a simplified exactly solvable model \cite{KS3}, where
we have been able to analyze the effects due to the absence of self -- 
averaging \cite{KS1,KS3,Kuch}.

The aim of the present paper is to analyze the main properties of 
superconducting state (for different types of pairing), appearing on the
``background'' of the pseudogap of ``dielectric'' nature in more realistic 
model of ``hot spots'' on the Fermi surface. Here we shall limit ourselves to
the region close to superconducting temperature $T_c$ and perform an
analysis based on microscopic derivation of Ginzburg -- Landau expansion,
assuming the self -- averaging nature of superconducting order parameter and
generalizing similar approach used earlier for ``hot patches'' model in
Ref. \cite{KS2}.
 
\section{``Hot -- Spots'' model and pairing interaction.}

\begin{figure}
\epsfxsize=16cm
\epsfysize=20cm
\epsfbox{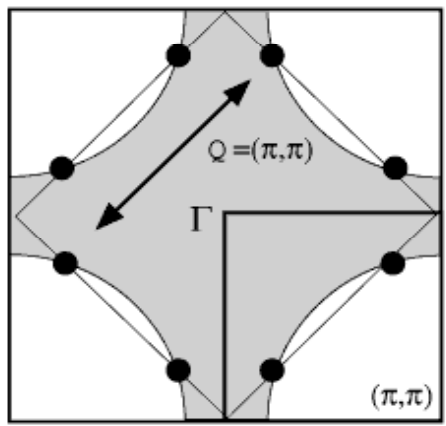}
\caption{Fermi surface with ``hot spots'' connected by scattering vector 
of the order of ${\bf Q}=(\frac{\pi}{a},\frac{\pi}{a})$.}
\label{hspts}
\end{figure}

In the model of ``nearly antiferromagnetic'' Fermi -- liquid, which is 
actively used to describe the microscopic nature of high -- temperature
superconductivity \cite{MBP,MP}, it is usually assumed that the effective
interaction of electrons with spin fluctuations of antiferromagnetic (AFM(SDW))
short -- range order is of the following form:
\begin{equation}
V_{eff}({\bf q},\omega)=
\frac{g^2\xi^2}{1+\xi^2({\bf q-Q})^2-i\frac{\omega}{\omega_{sf}}}
\label{V}
\end{equation}
where $g$ -- is some interaction constant, 
$\xi$ -- correlation length of spin fluctuations,
${\bf Q}=(\pi/a,\pi/a)$ -- vector of antiferromagnetic ordering in dielectric
phase, $\omega_{sf}$ -- characteristic frequency of spin fluctuations.
Both dynamic spin susceptibility and effective interaction
(\ref{V}) are peaked in the region of ${\bf q}\sim {\bf Q})$, which leads to
the appearance of ``two types'' of quasiparticles -- ``hot'' one, with momenta
in the vicinity of ``hot spots'' on the Fermi surface (Fig. \ref{hspts}), and
``cold'' one, with momenta close to the parts of the Fermi surface, surrounding
diagonals of the Brillouin zone \cite{Sch}. This is due to the fact, that 
quasiparticles from the vicinity of the ``hot spots'' are strongly scattered 
with the momentum transfer of the order of ${\bf Q}$, due to interaction
with spin fluctuations (\ref{V}), while for quasiparticles with momenta far
from these ``hot spots'' this interaction is relatively weak.

For high enough temperatures $\pi T\gg \omega_{sf}$,
we can neglect spin dynamics \cite{Sch}, limiting ourselves to static
approximation in (\ref{V}). Considerable simplification, allowing to analyze
higher -- order contributions, can be achieved by substitution of (\ref{V}) by
model -- like static interaction of the following form \cite{KS}:  
\begin{equation} 
V_{eff}({\bf q})=W^2\frac{2\xi^{-1}}{\xi^{-2}+(q_x-Q_x)^2}
\frac{2\xi^{-1}}{\xi^{-2}+(q_y-Q_y)^2}
\label{Veff}
\end{equation}
where $W$ is an effective parameter with dimension of energy. In the following,
as in Refs. \cite{Sch,KS}, we consider parameters $W$ as $\xi$
phenomenological (to be determined from the experiment). Anyhow, 
Eq. (\ref{Veff}) is qualitatively quite similar to the static limit of
(\ref{V}) and almost indistinguishable from it in most interesting region of
$|{\bf q-Q}|<\xi^{-1}$, determining scattering in the vicinity of ``hot spots''.

The spectrum of ``bare'' (free) quasiparticles can be taken as \cite{Sch}:
\begin{equation}
\xi_{\bf p}=-2t(\cos p_xa+\cos p_ya)-4t^{'}\cos p_xa\cos p_ya - \mu
\label{spectr}
\end{equation}
where $t$ is the transfer integral between nearest neighbors, while $t'$ is
the transfer integral between second nearest neighbors on the square lattice, 
$a$ is the lattice constant, $\mu$ -- chemical potential. This expression
gives rather good approximation to the results of band structure calculations 
of real HTSC -- system, e.g. for $YBa_2Cu_3O_{6+\delta}$ we have $t=0.25eV$,\ 
$t'=-0.45t$ \cite{Sch}. Chemical potential $\mu$ is fixed by concentration of
carriers. 

The least justified is an assumption of the static nature of fluctuations,
which can be valid only for rather high temperatures \cite{Sch,KS}. For low
temperatures, particularly in superconducting phase, spin dynamics can become
quite important, e.g. for microscopics of Cooper pairing in the model of
``nearly antiferromagnetic'' Fermi liquid \cite{MBP,MP}. However, we assume
here that the static approximation is sufficient for qualitative understanding
of the influence of pseudogap formation upon superconductivity, which will be
modelled within phenomenological BCS -- like approach.

In the limit of infinite correlation length
$\xi\to\infty$ this model acquires an exact solution \cite{C74}. For finite 
$\xi$ we can construct ``nearly exact'' solution \cite{KS}, generalizing
the one -- dimensional approach, proposed in Ref. \cite{C79}. Then we can
(approximately) sum the whole diagrammatic series for the one -- particle
electronic Green's function.

For the contribution of an arbitrary diagram for electronic self -- energy,
in the $N$-th order over the interaction (\ref{Veff}), we write down the
following {\em Ansatz} \cite{KS,C79}:
\begin{eqnarray}
\Sigma^{(N)}(\varepsilon_n{\bf p})=W^{2N}\prod_{j=1}^{2N-1}G_{0k_j}
(\varepsilon_n{\bf p}), \nonumber\\
G_{0k_j}(\varepsilon_n{\bf p})=\frac{1}{i\varepsilon_n-\xi_{k_j}({\bf p})+
ik_jv_{k_j}\kappa}
\label{Ansatz} 
\end{eqnarray}
where $\kappa=\xi^{-1}$, $k_j$ -- is the number of interaction lines,
surrounding the $j$-th (from the beginning) electronic line in a given
diagram, $\varepsilon_n=2\pi T(n+1/2)$ (assuming $\varepsilon_n>0$).
\begin{equation}
\xi_k({\bf p})=\left\{\begin{array}{ll}
\xi_{{{\bf p}}+{\bf Q}} & \mbox{for odd $k$} \\
\xi_{{\bf p}} & \mbox{for even $k$}
\end{array} \right.
\label{xik}
\end{equation}
\begin{equation}
v_k=\left\{\begin{array}{ll}
|v_x({\bf p}+{\bf Q})|+|v_y({\bf p}+{\bf Q})| & \mbox{for odd $k$} \\
|v_x({\bf p})|+|v_y({\bf p})| & \mbox{for even $k$}
\end{array} \right.
\label{Vk}
\end{equation}
where ${\bf v}({\bf p})=\frac{\partial\xi_{\bf p}}{\partial {\bf p}}$ --
velocity of a ``bare'' (free) quasiparticle.

In this approximation the contribution of an arbitrary diagram is determined,
in fact, by the set of integers $k_j$. Any diagram with intersection of 
interaction lines is actually equal to some diagram of the same order without
intersections and the contribution of all diagrams with intersections can be
accounted with the help of combinatorial factors $s(k_j)$ attributed to
interaction lines on diagrams without intersections \cite{C79,KS,Sch}. 

Combinatorial factor:
\begin{equation}
s(k)=k
\label{vcomm}
\end{equation}
in the case of commensurate fluctuations with
${\bf Q}=(\pi/a,\pi/a)$ \cite{C79}, if we neglect their spin structure 
\cite{Sch}) (i.e. limiting ourselves to CDW -- type fluctuations). Taking into
account the spin structure of interaction within the model of 
``nearly antiferromagnetic'' Fermi -- liquid (spin -- fermion model of 
Ref. \cite{Sch}), we obtain more complicated combinatorics of diagrams.
In particular, spin -- conserving scattering leads to the formally commensurate
combinatorics, while the spin -- flip scattering is described by diagrams
of incommensurate case (``charged'' random field in terms of Ref. \cite{Sch}).
As a result the {\em Ansatz} (\ref{Ansatz}) for one -- particle Green's function
is conserved, but the combinatorial factor $s(k)$ takes the form \cite{Sch}:
\begin{equation} 
s(k)=\left\{\begin{array}{cc}
\frac{k+2}{3} & \mbox{for odd $k$} \\
\frac{k}{3} & \mbox{for even $k$}
\end{array} \right.
\label{vspin}
\end{equation}
However, in the case of two -- particle processes and in the analysis of the
pseudogap influence on superconductivity, the use of spin -- fermion model
leads to significant complications. In particular, in this model there is a 
significant difference between vertexes, corresponding to spin dependent and
charge interactions, and combinatorics of diagrams for spin dependent vertex is
different from that of (\ref{vspin}) \cite{Sch}.

In spin -- fermion model \cite{Sch} the spin dependent part of interaction is
usually described by isotropic Heisenberg model. If for this interaction we
assume the Ising like form, only spin -- conserving processes remain and
commensurate combinatorics of diagrams (\ref{vcomm}) remains valid both for
one -- particle Green's function and for spin and charge vertexes. For this
reason, in the present work we shall limit ourselves to the analysis of 
commensurate (\ref{vcomm}) ``Ising like'' spin -- fluctuations only (AFM, SDW),
as well as commensurate charge fluctuations (CDW). Details concerning 
incommensurate fluctuations of CDW -- type can be found in 
Refs. \cite{KS,C74,C79}.

As a result we obtain the following recurrence procedure for the
one -- particle Green's function $G(\varepsilon_n{\bf p})$
(continuous fraction representation) \cite{C79,KS,Sch}:
\begin{equation}
G_k(\varepsilon_n{\bf p})=\frac{1}{i\varepsilon_n-\xi_{k}({\bf p})+
ikv_k\kappa-W^2s(k+1)G_{k+1}(\varepsilon_n{\bf p})}
\label{Gk}
\end{equation}
and ``physical'' Green's function is determined as
$G(\varepsilon_n{\bf p})\equiv G_0(\varepsilon_n{\bf p})$.

{\em Ansatz} (\ref{Ansatz}) for the contribution of an arbitrary diagram of
$N$-th order is not exact in general case \cite{KS}. However, for two --
dimensional system we were able to show that for certain topologies of the
Fermi surface (\ref{Ansatz}) is actually an exact representation \cite{KS}, 
in other cases it can be shown \cite{KS}, that this approximation overestimates
(in some sense) the effects of the finite values of correlation length
$\xi$ for the given order of perturbation theory. For one -- dimensional case,
when this problem is most serious \cite{KS}, the values of the density of
states found with the help of (\ref{Ansatz}) in the case of incommensurate
fluctuations are in almost ideal quantitative correspondence \cite{C00} with 
the results of an exact numerical modelling of this problem in Refs.
\cite{Kop,Mill}. In the limit of $\xi\to\infty$ {\em Ansatz} (\ref{Ansatz}) 
reduces to an exact solution of Ref. \cite{C74}, while in the limit of
$\xi\to 0$ and fixed value of $W$ it reduces to the trivial case of free
electrons.

This model describes non Fermi -- liquid like spectral density in the vicinity
of ``hot spots'' on the Fermi surface and ``smooth'' pseudogap in the density
of states \cite{Sch,KS}. 

To analyze superconductivity in such a system with well developed fluctuations
of short -- range order we shall assume that superconducting pairing is
determined by attractive BCS -- like interaction (between electrons with 
opposite spins) of the following simplest possible form:
\begin{equation} 
V_{sc}({\bf p,p'})=-Ve({\bf p})e({\bf p'}), 
\label{VV} 
\end{equation} 
where $e({\bf p})$ is given by:  
\begin{equation}
e({\bf p})=
\left\{
\begin{array}{ll}
1 & (\mbox{ $s$-wave pairing})\\ 
cos(p_xa)-cos(p_ya) & (\mbox{ $d_{x^2-y^2}$-wave pairing})\\
sin(p_xa)sin(p_ya) & (\mbox{ $d_{xy}$-wave pairing})\\
cos(p_xa)+cos(p_ya) & (\mbox{ anisotropic $s$-wave pairing})
\end{array}.
\right.
\label{ephi}
\end{equation}
Interaction constant $V$ is assumed, as usual, to be non zero in some interval
of the width of $2\omega_c$ around the Fermi level ($\omega_c$ -- 
characteristic frequency of quanta responsible for attractive interaction
of electrons). Then in general case the superconducting gap is anisotropic 
and given by: $\Delta({\bf p})=\Delta e({\bf p})$.

In Refs. \cite{SP,KS1} we have analyzed the peculiarities of superconducting
state in an exactly solvable model of the pseudogap state, induced by
short -- range order fluctuations with infinite correlation length
($\xi\to\infty$). In particular, in Ref. \cite{KS1} it was shown that these
fluctuations may lead to strong fluctuations of superconducting order
parameter (energy gap $\Delta$), which break the standard assumption of the
self -- averaging nature of the gap \cite{Gor,Genn,Scloc}, which allows to
perform independent averaging (over the random configurations of static
fluctuations of short -- range order) of superconducting order parameter
$\Delta$ and different combinations of electronic Green's functions, entering
the main equations. Usual argument for the possibility of such an independent
averaging goes as follows \cite{Gor,Scloc}: the value of $\Delta$ significantly
changes on the length scale of the order of coherence length
$\xi_0\sim v_F/\Delta_0$ of BCS theory, while the Green's functions
oscillate on much shorter length scales of the order of interatomic spacing.
Naturally, this assumption becomes invalid with the appearance (in electronic
system) of a new characteristic length scale $\xi\to\infty$. However, when
this correlation length of short -- range order $\xi\ll\xi_0$ 
(i.e. when fluctuations are correlated on the length scale shorter than the
characteristic size of Cooper pairs), the assumption of self -- averaging of 
$\Delta$ is apparently still valid, being broken only in case of $\xi>\xi_0$
\footnote{The absence of self -- averaging of superconducting gap even for the
case of $\xi<\xi_0$, obtained in Ref. \cite{KS3}, is apparently due to rather 
special model of short -- range order used in this paper.}. Thus, below we 
perform all the analysis under the usual assumption of self -- averaging 
energy gap over the fluctuations of short -- range order, which allows us to
use standard methods of the theory of disordered superconductors (mean field
approximation in terms of Ref. \cite{KS1}).

\section{Cooper instability.\ Recurrence procedure for the vertex part.}

It is well known that critical temperature of superconducting transition
can be determined from the equation for Cooper instability of normal phase:
\begin{equation}
1-V\chi(0;T)=0
\label{coopinst}
\end{equation}
where the generalized Cooper susceptibility is determined by diagram shown
in Fig. \ref{loop} and equal to:
\begin{equation}
\chi({\bf q};T)=-T\sum_{\varepsilon_n}\sum_{\bf p,p'}e({\bf p})e({\bf p'})
\Phi_{\bf p,p'}(\varepsilon_n,-\varepsilon_n,{\bf q})
\label{chi}
\end{equation}
where $\Phi_{\bf p,p'}(\varepsilon_n,-\varepsilon_n,q)$ is two -- particle 
Green's function in Cooper channel, taking into account scattering by
fluctuations of short -- range order.

Let us consider first the case of charge (CDW) fluctuations, when electron --
fluctuation interaction is spin independent. In case of isotropic $s$ and 
$d_{xy}$-wave pairing superconducting gap does not change after scattering 
by $\bf Q$, i.e. $e({\bf p}+{\bf Q})=e({\bf p})$ and 
$e({\bf p'})\approx e({\bf p})$. In case of anisotropic $s$ and $d_{x^2-y^2}$ 
pairings superconducting gap changes sign after scattering by $\bf Q$, 
i.e. $e({\bf p}+{\bf Q})=-e({\bf p})$, thus 
$e({\bf p'})\approx e({\bf p})$ for ${\bf p'}\approx{\bf p}$  and
$e({\bf p'})\approx -e({\bf p})$ for ${\bf p'}\approx{\bf p+Q}$. 
Thus, for diagrams with even number of interaction lines, connecting the upper
($\varepsilon_n$) and lower ($-\varepsilon_n$) electronic lines in Fig. 
\ref{loop}, we have ${\bf p'}\approx{\bf p}$ and the appropriate contribution
to susceptibility is the same for both isotropic $s$ and $d_{xy}$-wave pairing. 
However, for diagrams with odd number of such interaction lines we obtain 
contributions differing by sign. This sign change can be accounted by changing
the sign of interaction line, connecting the upper and lower lines in the 
loop in Fig. \ref{loop}. Then for the generalized susceptibility we obtain:
\begin{equation}
\chi({\bf q};T)=-T\sum_{\varepsilon_n}\sum_{\bf p}G(\varepsilon_n{\bf p+q})
G(-\varepsilon_n,-{\bf p})e^2({\bf p})
\Gamma^{\pm}(\varepsilon_n,-\varepsilon_n,{\bf q})
\label{chiq}
\end{equation}
where $\Gamma^{\pm}(\varepsilon_n,-\varepsilon_n,{\bf q})$ is 
``triangular'' vertex, describing electron interaction with fluctuations of
short -- range order, while the superscript $\pm$ denotes the abovementioned
change of signs for interaction lines connecting the upper and lower
electronic lines.

Consider now the case of scattering by spin (AFM(SDW)) fluctuations. In this
case interaction line, describing the longitudinal $S^z$ part of interaction,
surrounding the vertex changing spin direction, should be attributed an extra
factor of ($-1$) \cite{Sch}. Because of this factor, in case of electron 
interaction with spin fluctuations contributions to generalized susceptibility 
for different types of pairing considered above just ``change places''
\footnote{This is due to the fact of spin projections change in the vertex,
describing ``interaction'' with superconducting gap (restricting to the 
case of singlet pairing).} and susceptibility for the case of
isotropic $s$ and $d_{xy}$-wave pairing is determined by  ``triangular'' 
vertex $\Gamma^-$, while for anisotropic $s$ and $d_{x^2-y^2}$-wave case by 
``triangular'' vertex $\Gamma^+$.

Thus we come to the problem of calculation of ``triangular'' vertex parts, 
describing interaction with ``dielectric'' (pseudogap) fluctuations.
For one -- dimensional case and a similar problem (for real frequencies, 
$T=0$) the appropriate recursion procedure was formulated, for the first time,
in Ref. \cite{C91}. For a two -- dimensional model of the pseudogap with
``hot spots'' on the Fermi surface the generalization of this procedure
was performed in Ref. \cite{SS} and used to calculate optical conductivity.
All the details of appropriate derivation can also be found there.
The generalization to the case of Matsubara frequencies, of interest to us
here, can be made directly. Below, for definiteness we assume 
$\varepsilon_n>0$. Finally, for ``triangular'' vertex we obtain the recurrence
procedure, described by diagrams of Fig. \ref{recvertx}
(where the wavy line denotes interaction with pseudogap fluctuations),
and having the following analytic form:
\begin{eqnarray}
\Gamma^{\pm}_{k-1}(\varepsilon_n,-\varepsilon_n,{\bf q})=\nonumber\\
=1 \pm W^2s(k)G_k\bar G_k
\Biggl\{1+\frac{2ikv_k\kappa}{2i\varepsilon_n-{\bf v}_k{\bf q}-W^2s(k+1)
(G_{k+1}-\bar G_{k+1})}\Biggr\}\Gamma^{\pm}_{k}(\varepsilon_n,-\varepsilon_n,
{\bf q})  
\label{Gamma}
\end{eqnarray}
where $G_k=G_k(\varepsilon_n{\bf p+q})$ and 
$\bar G_k=G_k(-\varepsilon_n,-{\bf p})$ are calculated according to
(\ref{Gk}), $v_k$ is defined by (\ref{Vk}), and ${\bf v}_k$ are given by:
\begin{equation}
{\bf v}_k=\left\{\begin{array}{ll}
{\bf v}({\bf p}+{\bf Q}) & \mbox{for odd $k$} \\
{\bf v}({\bf p}) & \mbox{for even $k$}
\end{array} \right.
\label{bfVk}
\end{equation}
``Physical'' vertex is defined as 
$\Gamma^{\pm}(\varepsilon_n,-\varepsilon_n,{\bf q})\equiv 
\Gamma^{\pm}_{0}(\varepsilon_n,-\varepsilon_n,{\bf q})$.

To determine $T_c$ we need vertices at ${\bf q}=0$. Then $\bar G_k=G^*_k$ 
and vertex parts $\Gamma^+_k$ and $\Gamma^-_k$ become real significantly
simplifying our procedures (\ref{Gamma}). For $ImG_k$ and $ReG_k$ we obtain
the following system of recurrence equations:
\begin{eqnarray}
ImG_k=-\frac{\varepsilon_n+kv_k\kappa-W^2s(k+1)ImG_{k+1}}{D_k}\nonumber\\
ReG_k=-\frac{\xi_k({\bf p}) +W^2s(k+1)ReG_{k+1}}{D_k}
\label{JR}
\end{eqnarray}
where $D_k=(\xi_k({\bf p})+W^2s(k+1)ReG_{k+1})^2+
(\varepsilon_n+kv_k\kappa-W^2s(k+1)ImG_{k+1})^2$,
and vertex parts at ${\bf q}=0$ are determined by:
\begin{equation}
\Gamma^{\pm}_{k-1}=1\mp W^2s(k)\frac{ImG_k}
{\varepsilon_n-W^2s(k+1)ImG_{k+1}}\Gamma^{\pm}_{k}
\label{Gammk}
\end{equation}

Going to numerical calculations we have to define characteristic energy
(temperature) scale associated with superconductivity in the absence of
pseudogap fluctuations ($W=0$). In this case the equation for superconducting
critical temperature $T_{c0}$ becomes standard BCS -- like (for the general
case of anisotropic pairing):
\begin{equation}
1=\frac{2VT}{\pi^2}\sum_{n=0}^{\bar m}\int_{0}^{\pi}dp_x\int_{0}^{\pi}dp_y
\frac{e^2({\bf p})}{\xi^2_{\bf p}+\varepsilon^2_n}
\label{TcBCS}
\end{equation}
where $\bar m=[\frac{\omega_c}{2\pi T_{c0}}]$ is the dimensionless cutoff for
the sum over Matsubara frequencies. All calculations were performed for typical
quasiparticle spectrum in HTSC given by (\ref{spectr}) with $\mu=-1.3t$ and
$t'/t=-0.4$. Choosing, rather arbitrarily, the value of $\omega_c=0.4t$ and 
$T_{c0}=0.01t$ we can easily determine the appropriate values of pairing
constant $V$ in (\ref{TcBCS}), giving this value of $T_{c0}$ for different
types of pairing listed in (\ref{ephi}). In particular, for the usual
isotropic $s$-wave pairing we obtain $\frac{V}{ta^2}=1$, while for 
$d_{x^2-y^2}$ pairing we get $\frac{V}{ta^2}=0.55$. For other types of
pairing from (\ref{ephi}) the values of pairing constant for this choice of
parameters are found to be unrealistically large and we shall not quote the 
the results of numerical calculations for these symmetries
\footnote{Of course, our description, based on BCS equations of weak coupling
theory, is not pretending to be realistic also for the cases of $s$-wave and
$d_{x^2-y^2}$-wave pairing. We only need to define characteristic scale of
$T_{c0}$ and express all temperatures below in units of this temperature,
assuming certain universality of all dependences with respect to this scale.}.

Typical results of numerical calculations of superconducting transition
temperature $T_c$ for the system with pseudogap obtained using our recursion
relations directly from (\ref{coopinst}) are shown in 
Figs. \ref{Tc+},\ref{Tc-}. We can see that in all cases pseudogap 
(``dielectric'') fluctuations lead to significant reduction of superconducting
transition temperature. This reduction stronger in case of $d_{x^2-y^2}$-wave 
pairing than for isotropic $s$-wave case. At the same time, reduction of the
correlation length $\xi$ (growth of $\kappa$) of pseudogap fluctuations leads
to the growth of $T_c$. These results are similar to those obtained earlier
in the ``hot patches'' model \cite{SP,KS2}. However, significant qualitative
differences also appear. From Fig. \ref{Tc+} it is seen, that for the case of
isotropic $s$-wave pairing and scattering by charge (CDW) fluctuations, as well
as for $d_{x^2-y^2}$ pairing and scattering by spin (AFM(SDW)) fluctuations 
\footnote{This last case is apparently realized in copper oxides.}
(i.e. in cases when the upper sign is ``operational'' in Eqs. (\ref{Gamma}) and 
(\ref{Gammk}), leading to recursion procedure for the vertex with same signs)
there appears characteristic plateau in the dependence of $T_c$ on the width
of the pseudogap $W$ in the region of $W<10T_{c0}$, while significant
suppression of $T_c$ takes place on the scale of $W\sim 50 T_{c0}$. 
Qualitative differences appear also for the case of $s$-wave pairing and
scattering by spin (AFM(CDW)) fluctuations, as well as for the case of
$d_{x^2-y^2}$ pairing and scattering by charge fluctuations.
From Fig. \ref{Tc-} we can see that in this case (when lower sign is 
``operational'' in Eqs. (\ref{Gamma}) and (\ref{Gammk}), i.e. when we have 
recursion procedure for the vertex with alternating signs) reduction of $T_c$ 
is an order of magnitude faster. 
In the case of $d_{x^2-y_2}$ pairing, for the values of $W/T_{c0}$ corresponding
to almost complete suppression of superconductivity, our numerical 
procedures become unreliable. In particular, we can observe here 
characteristic non single valued dependence of $T_c$ on $W$, which may 
signify the existence of a narrow region of phase diagram with ``reentrant'' 
superconductivity\footnote{Our calculations show that manifestations of such 
behavior of $T_c$ become stronger for the case of scattering by 
incommensurate pseudogap fluctuations.}. 
This behavior of $T_c$ resembles similar dependences,
appearing in superconductors with Kondo impurities \cite{Zitt}. Alternative
possibility is the appearance here of the region, where superconducting
transition becomes first order, similarly to situation in superconductors
with strong paramagnetic effect in external magnetic field\cite{Sarma}. 
However, it should be stressed that our calculations mostly show the 
probable appearance of some critical value of $W/T_{c0}$, corresponding to
the complete suppression of superconductivity.
In any case, the observed behavior requires special studies in the 
future and below we only give the results, corresponding to the region of
single -- valued dependences of $T_c$.

\section{Ginzburg -- Landau expansion.}

In Ref. \cite{SP} Ginzburg -- Landau expansion was derived for an exactly
solvable model of the pseudogap with infinite correlation length of short --
range order fluctuations. In Ref. \cite{KS2} these results were extended for
the case of finite correlation lengths. In these papers the analysis was, in 
fact, done only for the case of charge fluctuations and simplified model
of pseudogap state, based on the picture of ``hot'' (flat) parts (patches)
on the Fermi surface. Also in this model the signs of superconducting gap
after the transfer by vector ${\bf Q}$ were assumed to be the same, both for
$s$-wave and $d$-wave pairings \cite{KS2}. Here we shall make appropriate
generalization for the present more realistic model of ``hot spots'' on 
the Fermi surface. 

Ginzburg -- Landau expansion for the difference of free energy density of
superconducting and normal states can be written in the usual form:
\begin{equation}
F_{s}-F_{n}=A|\Delta_{\bf q}|^2
+q^2 C|\Delta_{\bf q}|^2+\frac{B}{2}|\Delta_{\bf q}|^4,
\label{GiLa}
\end{equation}
where $\Delta_{\bf q}$ is the amplitude of the Fourier component of the order
parameter, which can be written for different types of pairing as:
$\Delta({\bf p},{\bf q})=\Delta_qe({\bf p})$. In fact (ref{GiLa}) is
determined by diagrams of loop expansion of free energy in the field of
random fluctuation of the order parameter (denoted by dashed lines) with small
wave vector ${\bf q}$ \cite{SP}, shown in Fig. \ref{GL}. 

Coefficients of Ginzburg -- Landau expansion can be conveniently expressed as:
\begin{equation}
A=A_0K_A;\qquad   C=C_0K_C;\qquad    B=B_0K_B,
\label{ACD}
\end{equation}
where $A_0$, $C_0$ and $B_0$ denote expressions of these coefficients
(derived in the Appendix) in the absence of pseudogap fluctuations 
($W=0$) for the case of an arbitrary quasiparticle spectrum $\xi_p$ and
different types of pairing: 
\begin{eqnarray} 
&&A_0=N_0(0)\frac{T-T_{c}}{T_{c}}<e^2({\bf p})>;\quad 
C_0=N_0(0)\frac{7\zeta(3)}{32\pi^{2}T_c^2}<|{\bf v}({\bf p})|^2e^2({\bf p)}>; 
\nonumber\\ 
&&B_0=N_0(0)\frac{7\zeta(3)}{8\pi^{2}T_c^2}<e^4({\bf p})>,
\label{ACDf}
\end{eqnarray}
where angular brackets denote the usual averaging over the Fermi surface:
$<\ldots>=\frac{1}{N_0(0)}\sum_p\delta (\xi_{\bf p})\ldots$,\ 
and $N_0(0)$ is the density of states at the Fermi level for free electrons.

Now all peculiarities of the model under consideration, due to the appearance
of the pseudogap, are contained within dimensionless coefficients
$K_A$, $K_C$ and $K_B$. In the absence of pseudogap fluctuations all these
coefficients are equal to 1.

Coefficients $K_A$ and $K_C$, according to Fig. \ref{GL}(a) are completely
determined by the generalized Cooper susceptibility \cite{SP,KS2} 
$\chi({\bf q};T)$, shown in Fig. \ref{loop}:  
\begin{equation} 
K_A=\frac{\chi(0;T)-\chi(0;T_c)}{A_0}
\label{Ka}
\end{equation}
\begin{equation}
K_C=\lim_{q\to 0}\frac{\chi({\bf q};T_c)-\chi(0;T_c)}{q^2C_0}
\label{Kc}
\end{equation}
Generalized susceptibility, as was shown above, can be found from
(\ref{chiq}), where ``triangular'' vertices are determined by recurrence
procedures (\ref{Gamma}), allowing to perform direct numerical calculations of 
the coefficients $K_A$ and $K_C$. 

Situation with coefficient $B$ is, in general case, more complicated.
Significant simplifications arise if we limit ourselves (in the order
$|\Delta_q|^4$), as usual, to the case of $q=0$, and define the coefficient
$B$ by diagram shown in Fig. \ref{GL}(b). Then for coefficient $K_B$ we get:
\begin{equation}
K_B=\frac{T_c}{B_0}\sum_{\varepsilon_n}\sum_{\bf p}e^4({\bf p})
(G(\varepsilon_n{\bf p})G(-\varepsilon_n,-{\bf p}))^2
(\Gamma^{\pm}(\varepsilon_n,-\varepsilon_n,0))^4
\label{Kbs}
\end{equation}
It should be noted that Eq. (\ref{Kbs}) immediately leads to positively
defined coefficient $B$. This is clear from
$G(-\varepsilon_n,-{\bf p})=G^*(\varepsilon_n{\bf p})$, so that
$G(\varepsilon_n{\bf p})G(-\varepsilon_n,-{\bf p})$ is real and accordingly
$\Gamma^{\pm}(\varepsilon_n,-\varepsilon_n,0)$, defined by recurrence
procedure (\ref{Gammk}), is also real.

\section{Physical properties of superconductors in the pseudogap state.}

It is well known that Ginzburg -- Landau equations define two
characteristic lengths --- coherence length and penetration depth.
Coherence length at a given temperature $\xi(T)$ determines characteristic
length scale of inhomogeneities of the order parameter $\Delta$:
\begin{equation}
\xi^2(T)=-\frac{C}{A}.
\label{xii}
\end{equation}
In the absence of the pseudogap:
\begin{equation}
\xi_{BCS}^2(T)=-\frac{C_0}{A_0}
\label{xi}
\end{equation}
Thus, in our model:
\begin{equation}
\frac{\xi^2(T)}{\xi_{BCS}^2(T)}=\frac{K_C}{K_A}.
\label{xiii}
\end{equation}
For the penetration depth of magnetic field we have:
\begin{equation}
\lambda^2(T)=-\frac{c^2}{32\pi e^2}\frac{B}{AC}
\label{lam}
\end{equation}
Then, analogously to (\ref{xiii}), we obtain:
\begin{equation}
\frac{\lambda(T)}{\lambda_{BCS}(T)}=
\left(\frac{K_{B}}{K_{A}K_{C}}\right)^{1/2}.
\label{lm}
\end{equation}
Close to $T_{c}$ the upper critical field $H_{c2}$ is defined via Ginzburg --
Landau coefficients as:  
\begin{equation} 
H_{c2}=\frac{\phi_0}{2\pi\xi^2(T)}=-\frac{\phi_{0}}{2\pi}\frac{A}{C} ,
\label{Hc2} 
\end{equation} 
where $\phi_{0}=c\pi/|e|$ is the magnetic flux quantum. 
Then the derivative (slope) of the upper critical field close to $T_{c}$
is given by:  
\begin{equation} 
\left|\frac{dH_{c2}}{dT}\right|_{T_c}=
\frac{16\pi\phi_{0}<e^2({\bf p})>}
{7\zeta(3)<|{\bf v}({\bf p})|^2e^2(\bf p)>}T_{c}
\frac{K_A}{K_C}. 
\label{dHc2}
\end{equation}
Specific heat discontinuity at the transition point is defined as:
\begin{equation} 
(C_s-C_n)_{T_c}=\frac{T_c}{B}\left(\frac{A}{T-T_c}\right)^2,
\label{Cs}
\end{equation}
where $C_s$ and $C_n$ are specific heats of superconducting and normal states.
At temperature $T_{c0}$ (in the absence of the pseudogap, $W=0$):
\begin{equation}
(C_s-C_n)_{T_{c0}}=N(0)
\frac{8\pi^2T_{c0}<e^2({\bf p})>^2}{7\zeta(3)<e^4({\bf p})>}.
\label{CsCn}
\end{equation}
Then the relative specific heat discontinuity in our model can be written as:
\begin{equation}
\Delta C\equiv\frac{(C_s-C_n)_{T_c}}{(C_s-C_n)_{T_{c0}}}=
\frac{T_c}{T_{c0}}\frac{K_A^2}{K_B}.
\label{cscn}
\end{equation}
Numerical calculations of $K_A,\ K_B,\ K_C$ were performed for the same typical
parameters of the model as calculation of $T_c$ described above. 
The values of these coefficients, are not of great interest and we
drop these data\footnote{Typical dependences of these coefficients on the
parameter $W/T_{c0}$ are functions rather rapidly decreasing from 1 in the
region of existence of superconducting state.} presenting in 
Figs. \ref{xi+}--\ref{DC-} appropriate $W/T_{c0}$ -- dependences of physical 
characteristics, defined by Eqs. (\ref{xii}) -- (\ref{cscn}). 
In accord with typical behavior of $T_c$ described above, here we can also
observe two qualitatively different types of behavior cue to either
constant or alternating signs in recurrence equations for the vertex part
(upper or lower signs in (\ref{Gamma}), charge or spin fluctuations). 
Results of our calculations of physical characteristics for the first
type of behavior ($s$-wave pairing and scattering by charge (CDW) fluctuations,
$d_{x^2-y^2}$-wave pairing and scattering by spin (AFM(SDW)) fluctuations) 
are shown in Figs. \ref{xi+} -- \ref{DC+}. We can see that with the growth
of pseudogap width $W$ coherence length $\xi(T)$ is reduced, while penetration
depth $\lambda(T)$ grows, as compared to the appropriate values of BCS theory. 
Dependence of both characteristic lengths on parameter $\kappa a$ is rather
weak and in Figs. \ref{xi+}, \ref{lamb+} we present data for only one value
of $\kappa a=0.2$. The slope (derivative) of the upper critical field at
$T=T_c$ initially grows, then starts to diminish. Most characteristic is the 
suppression of specific heat discontinuity (as compared to BCS value) shown
in Fig. \ref{DC+}, which is in direct qualitative accord with experimental
data \cite{Lorm}. Note that for specific heat discontinuity we also have
noticeable plateau in the region of  $W/T_{c0}<10$, analogous to that
observed above in similar dependence of $T_{c}$.

Behavior of physical characteristics of superconductor for the case of
$s$-wave pairing and scattering by spin (AFM(CDW)) fluctuations as well as for
$d_{x^2-y^2}$ pairing and scattering by charge (CDW) fluctuations are shown
in Figs. \ref{Hc2-}, \ref{DC-}. Data on characteristic lengths are not given,
as both coherence length $\xi(T)$, and penetration depth $\lambda(T)$ are 
practically the same as in BCS theory in almost all region of existence of
superconducting state (except very narrow region where $T_c$ can be non
single valued and disappear, where both lengths are rapidly growing). 
As to the slope of an upper critical field and specific heat discontinuity
at superconducting transition, both are rapidly suppressed with the growth
of $W/T_{c0}$, up to the region of non single valued $T_c$ (the region of 
possible ``reentrant'' behavior or type I transition)
(cf. Fig. \ref{Hc2-}, \ref{DC-}), where both are characterized by non 
monotonous behavior.

\section{Conclusion}

In this paper we have analyzed peculiarities of superconducting state 
forming on the ``background'' of the pseudogap induced by electron scattering
by fluctuations of ``dielectric'' short -- range order in the model of
``hot spots'' on the Fermi surface. Our analysis is based on microscopic 
derivation of Ginzburg -- Landau expansion and takes into account all
high order contributions of perturbation theory for scattering by pseudogap
fluctuations. ``Condensed'' phase of such superconductor should be described
by appropriate Gor'kov's equations for a superconductor with the pseudogap
\cite{KS2}, which will be the task for the future work.

Our main conclusion is that superconductivity is suppressed by pseudogap
fluctuations of CDW, AFM(SDW) type with two qualitatively different model of
such suppression, related to either constant or alternating signs in recurrence
equations for the vertex part (upper or lower signs in (\ref{Gamma}, charge or
spin fluctuations). While the case of scattering by spin fluctuations and
pairing with $d_{x^2-y^2}$ -- symmetry is apparently realized in high --
temperature superconducting copper oxides, we are not aware of systems with
unusual behavior obtained above for the case of
$s$-wave pairing and scattering by spin (AFM(CDW)) fluctuations, as well as
for the case of $d_{x^2-y^2}$ pairing and scattering by charge (CDW)
fluctuations. Experimental search of systems with this behavior may be of
significant interest. 

The most important problem in the description of the pseudogap state in
real HTSC -- cuprates is an explanation of doping dependences of the main
physical characteristics. In our model, these dependences should be expressed
via corresponding doping dependences of the effective pseudogap width $W$ and
correlation length $\xi$. Unfortunately, these dependences are determined
from rather indirect experimental data and are poorly defined 
\cite{Tim,MS}\footnote{Concentartion dependence of $T_{c0}$ may also be quite
important, but in fact it is completely unknown.}. 
As a very crude statement, we can claim that correlation length $\xi$ does not
change significantly in rather wide region of doping concentrations, while
the pseudogap width $W$ drops linearly with concentration growth from the
values of the order of $10^3K$ close to the insulating phase, to the values
of the order of $T_c$ at optimal doping, becoming zero at slightly higher 
concentrations of current carriers (Cf. Fig.6 in the review paper \cite{MS},
which is based of Fig.4 of Ref. \cite{Lor}, with some compilation of
appropriate data for $YBCO$). Using such doping dependence we can immediately
reexpress $W$ -- dependences given above as dependences on concentration of
doping impurity. In the oversimplfied version of our model with an infinite
correlation length and Fermi surface with complete nesting, assuming also
linear doping dependence for $T_{c0}$, such calculations were given in a 
recent paper \cite{AC}. It was shown that even within this simplest approach
the qualitative form of the phase diagram of cuprates can be nicely reproduced.
However, the obvious crudeness of the model, as well as the absence of reliable
data on concentration dependences of $W$, $\xi$ and $T_{c0}$, do not allow
any serious ``improvement'' of these qualitative conclusions. 

Among shortcomings of the present model, besides many times mentioned
neglect of dynamics of short -- range order fluctuations and limitation to
Gaussian statistics of these fluctuations, we also note the simplified
analysis of spin structure of interaction, which assumed its Ising like nature.
It will be of great interest to perform similar analysis for the general
case of Heisenberg interaction.

This work is partly supported by RFBR grant 02-02-16031, as well as by the
program of fundamental research of Presidium of the RAS 
``Quantum macrophysics'' and program of the Division of Physical Sciences of 
the RAS ``Strongly correlated electrons in semiconductors, metals, 
superconductors and magnetic materials'', and by the research project of
the Russian Ministry of Science and Industries ``The studies of collective and
quantum effects in condensed matter''.

\newpage

\appendix

\section{Ginzburg -- Landau coefficients for anisotropic superconductor
in the absence of the pseudogap.}

\begin{figure}
\epsfxsize=14cm
\epsfysize=20cm
\epsfbox{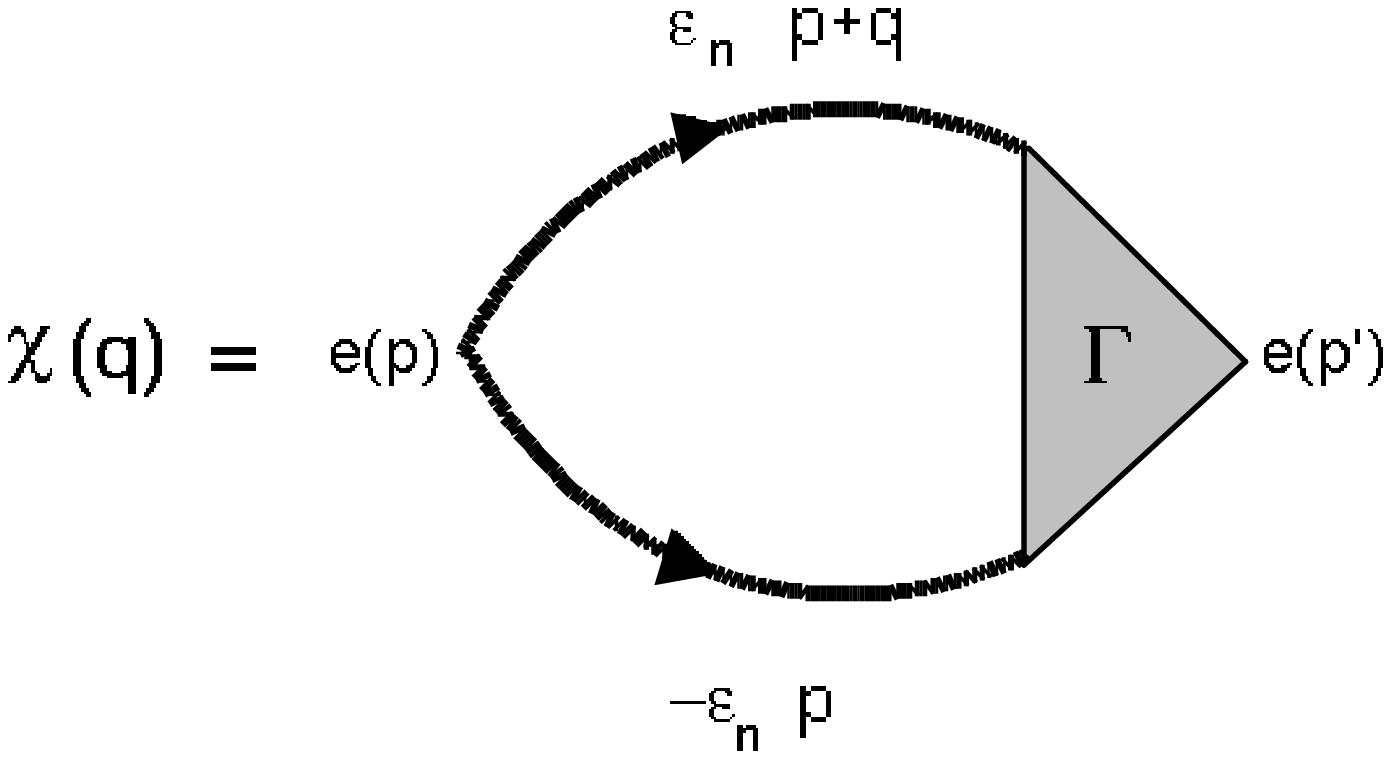}
\caption{Diagaram for the generalized susceptibility 
$\chi ({\bf q})$ in Cooper channel.}
\label{loop}
\end{figure}

\begin{figure}
\epsfxsize=14cm
\epsfysize=20cm
\epsfbox{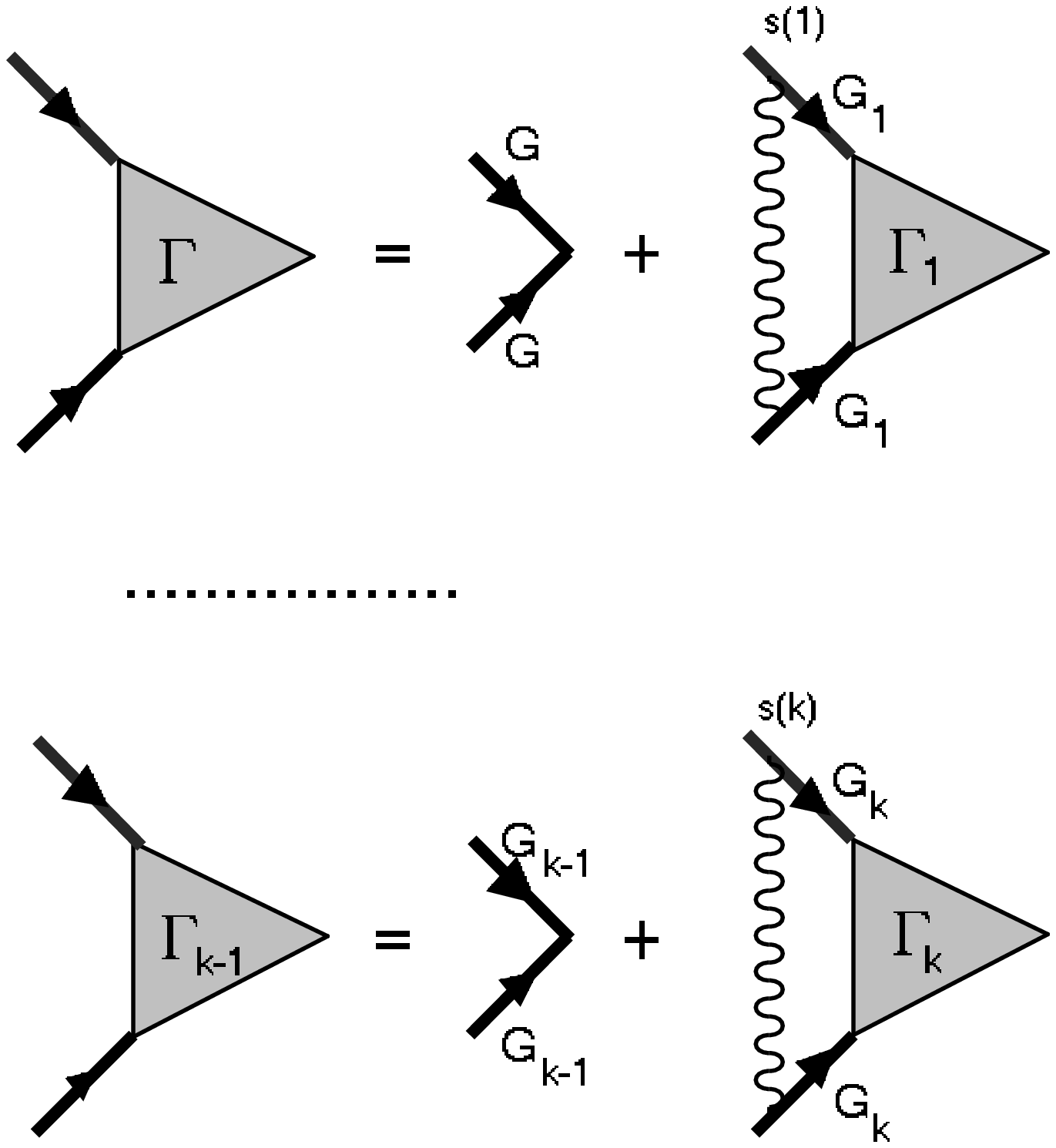}
\caption{Recurrence equations for the vertex part.}
\label{recvertx}
\end{figure}

\begin{figure}
\epsfxsize=14cm
\epsfysize=16cm
\epsfbox{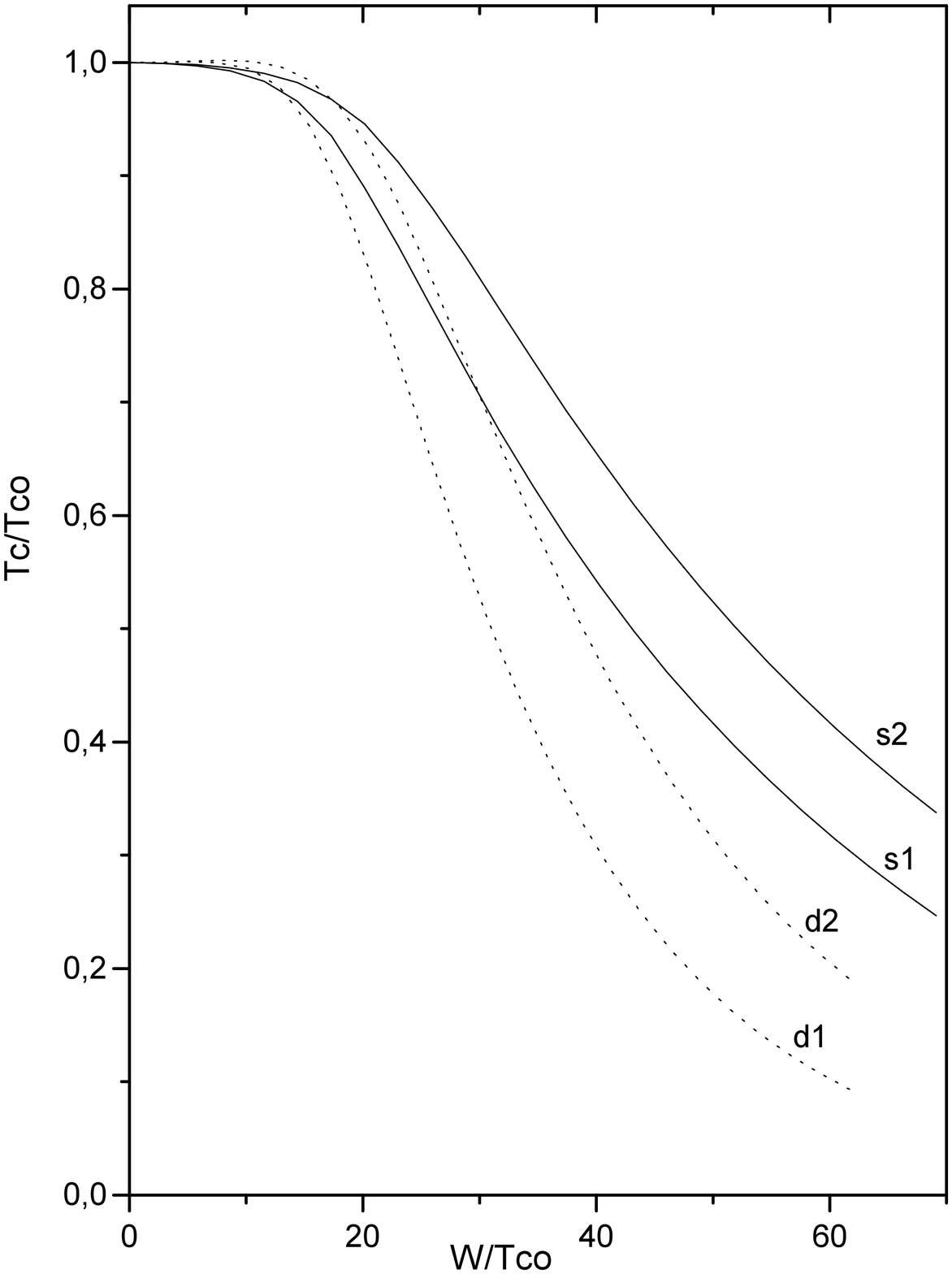}
\caption{Dependence of superconducting transition temperature $T_c/T_{c0}$ on
the effective width of the pseudogap $W/T_{c0}$ for $s$-wave pairing and
scattering by charge (CDW) fluctuations (curves s1 and s2) and for
$d_{x^2-y^2}$-wave pairing and scattering by spin (AFM(SDW)) fluctuations
(curves d1 and d2). Data are given for the values of inverse correlation
length ${\kappa a}=0.2$ (s1 and d1) and ${\kappa a}=0.5$ (s2 and d2).}  
\label{Tc+} 
\end{figure}

\begin{figure}
\epsfxsize=14cm
\epsfysize=16cm
\epsfbox{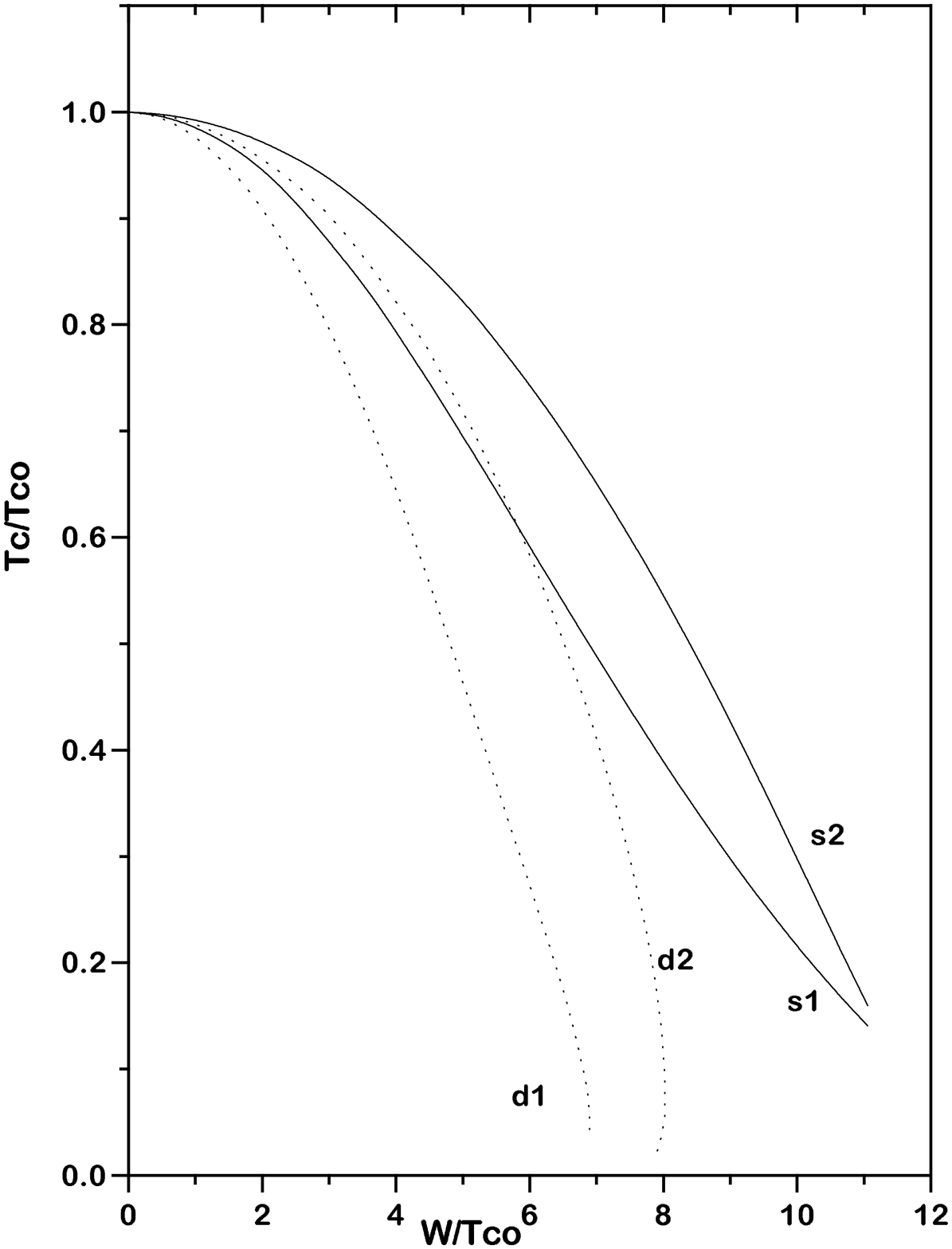}
\caption{Dependence of superconducting transition temperature $T_c/T_{c0}$ on
the effective width of the pseudogap $W/T_{c0}$ for $s$-wave pairing and
scattering by spin (AFM(SDW)) fluctuations (curves s1 and s2) and for
$d_{x^2-y^2}$-wave pairing and scattering by charge (CDW) fluctuations 
(curves d1 and d2). Data are given for the values of inverse correlation
length ${\kappa a}=0.2$ (s1 and d1) and ${\kappa a}=1.0$ (s2 and d2).}  
\label{Tc-} 
\end{figure}

In the absence of pseudogap fluctuations ($W=0$) the generalized Cooper
susceptibility define by diagram of Fig. \ref{loop} takes the following form:
\begin{equation}
\chi_0({\bf q};T)=-T\sum_{\varepsilon_n}\sum_{\bf p}e^2({\bf p})
\frac{1}{i\varepsilon_n-\xi_{{\bf p}+{\bf q}}}
\frac{1}{-i\varepsilon_n-\xi_{\bf p}}
\label{xioq}
\end{equation}
Then for susceptibility at ${\bf q}=0$, defining the coefficient $A_0$, 
we get: 
\begin{eqnarray}
\chi_0(0;T)=-T\sum_{\varepsilon_n}\sum_{\bf p}e^2({\bf p})
\frac{1}{\varepsilon_n^2+\xi_{\bf p}^2}=
-T\sum_{\varepsilon_n}\int_{-\infty}^{\infty}d\xi
\frac{1}{\varepsilon_n^2+\xi^2}
\sum_{\bf p}\delta(\xi-\xi_{\bf p})e^2({\bf p})\approx
\nonumber\\
\approx -N_0(0)T\sum_{\varepsilon_n}\int_{-\infty}^{\infty}d\xi
\frac{1}{\varepsilon_n^2+\xi^2}
\frac{\sum_{\bf p}\delta(\xi_{\bf p})e^2({\bf p})}{N_0(0)}=
\chi_{BCS}(0;T)<e^2({\bf p})>
\label{xioo}
\end{eqnarray}
where angular brackets denote averaging over the Fermi surface and we
introduced the standard susceptibility of BCS model for isotropic
$s$-wave pairing $\chi_{BCS}(0;T)$.

As a result, we obtain coefficient $A_0$ as:
\begin{equation}
A_0=\chi_0(0;T)-\chi_0(0;T_c)=A_{BCS}<e^2({\bf p})>
\label{Ao}
\end{equation}
where 
\begin{equation}
A_{BCS}=\chi_{BCS}(0;T)-\chi_{BCS}(0;T_c)=N_0(0)\frac{T-T_c}{T_c}
\label{Abcs}
\end{equation}
is the standard expression for coefficient $A$ for the case of isotropic
$s$-wave pairing.

Coefficient $C_0$ of Ginzburg -- Landau expansion is defined by susceptibility 
(\ref{xioq}) at small ${\bf q}$:
\begin{equation}
C_0=\lim_{q\to 0}\frac{\chi_0({\bf q};T_c)-\chi_0(0;T_c)}{q^2}
\label{Co}
\end{equation}
Expanding expression (\ref{xioq}) for $\chi_0({\bf q};T_c)$ in powers of $q$
we get:
\begin{equation}
\chi_0({\bf q};T_c)=\chi_0(0;T_c)+
T_c\sum_{\varepsilon_n}\sum_{\bf p}
\frac{3\varepsilon_n^2-\xi_{\bf p}^2}{4(\varepsilon_n^2+\xi_{\bf p}^2)^3}
e^2({\bf p})({\bf v}({\bf p}){\bf q})^2
\label{rxioq}
\end{equation}
so that coefficient $C_0$ is found to be:
\begin{eqnarray}
C_0=
T_c\sum_{\varepsilon_n}\int_{-\infty}^{\infty}d\xi
\frac{3\varepsilon_n^2-\xi^2}{4(\varepsilon_n^2+\xi^2)^3}
\sum_{\bf p}\delta(\xi-\xi_{\bf p})e^2({\bf p})|{\bf v}({\bf p})|^2cos^2(\phi)
\approx\nonumber\\
\approx T_c\sum_{\varepsilon_n}\int_{-\infty}^{\infty}d\xi
\frac{3\varepsilon_n^2-\xi^2}{4(\varepsilon_n^2+\xi^2)^3}
\sum_{\bf p}\delta(\xi_{\bf p})e^2({\bf p})|{\bf v}({\bf p})|^2cos^2(\phi)=
\nonumber\\
=N_0(0)\frac{7\zeta(3)}{16\pi^{2}T_c^2}
<e^2({\bf p})|{\bf v}({\bf p})|^2cos^2(\phi)>
\label{Coo}
\end{eqnarray}
where $\phi$ is an angle between vectors  ${\bf v}({\bf p})$ and ${\bf q}$, 
$\zeta(3)=\sum_{n=1}^{\infty}\frac{1}{n^3}\approx 1,202$

For square lattice both the Fermi surface and
$|{\bf v}({\bf p})|$ are symmetric with respect to rotation
over an angle $\frac{\pi}{2}$, the same symmetry is valid for
$e({\bf p})$ (for all types of pairing, mentioned above). 
Thus we easily find:
\begin{equation}
<e^2({\bf p})|{\bf v}({\bf p})|^2cos^2(\phi)>=
\frac{1}{2}<e^2({\bf p})|{\bf v}({\bf p})|^2(1+cos(2\phi))>=
\frac{1}{2}<e^2({\bf p})|{\bf v}({\bf p})|^2>
\label{simm}
\end{equation}
because $cos(2\phi)$ changes sign after the rotation of ${\bf p}$ by an angle
$\frac{\pi}{2}$. In fact, the direction of the velocity ${\bf v}({\bf p})$
after this rotation changes to a perpendicular one, accordingly
$cos(2\phi)\to -cos(2\phi)$. As a result we obtain for the coefficient
$C_0$ an isotropic expression:
\begin{equation}
C_0=N_0(0)\frac{7\zeta(3)}{32\pi^{2}T_c^2}<|{\bf v}({\bf p})|^2e^2({\bf p)}>,
\label{Coiz}
\end{equation}
which for the case of isotropic $s$-wave pairing and spherical Fermi surface
becomes the standard one:
\begin{equation}
C_{BCS}=N_0(0)\frac{7\zeta(3)v_F^2}{32\pi^{2}T_c^2} 
\label{Cbcs}
\end{equation}

\begin{figure}
\epsfxsize=14cm
\epsfysize=20cm
\epsfbox{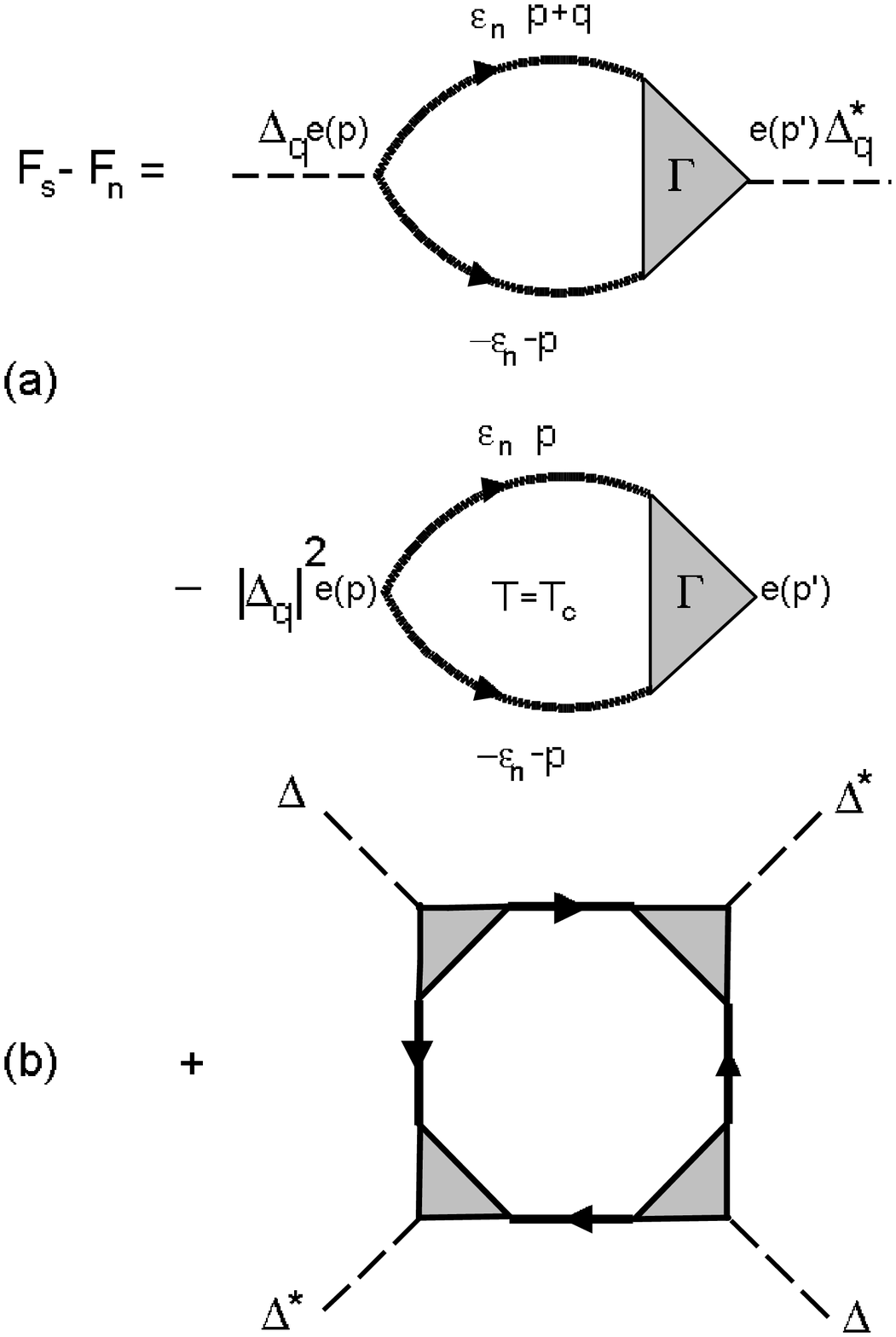}
\caption{Diagrammatic representation of Ginzburg -- Landau expansion.}
\label{GL}
\end{figure}

\begin{figure}
\epsfxsize=14cm
\epsfysize=16cm
\epsfbox{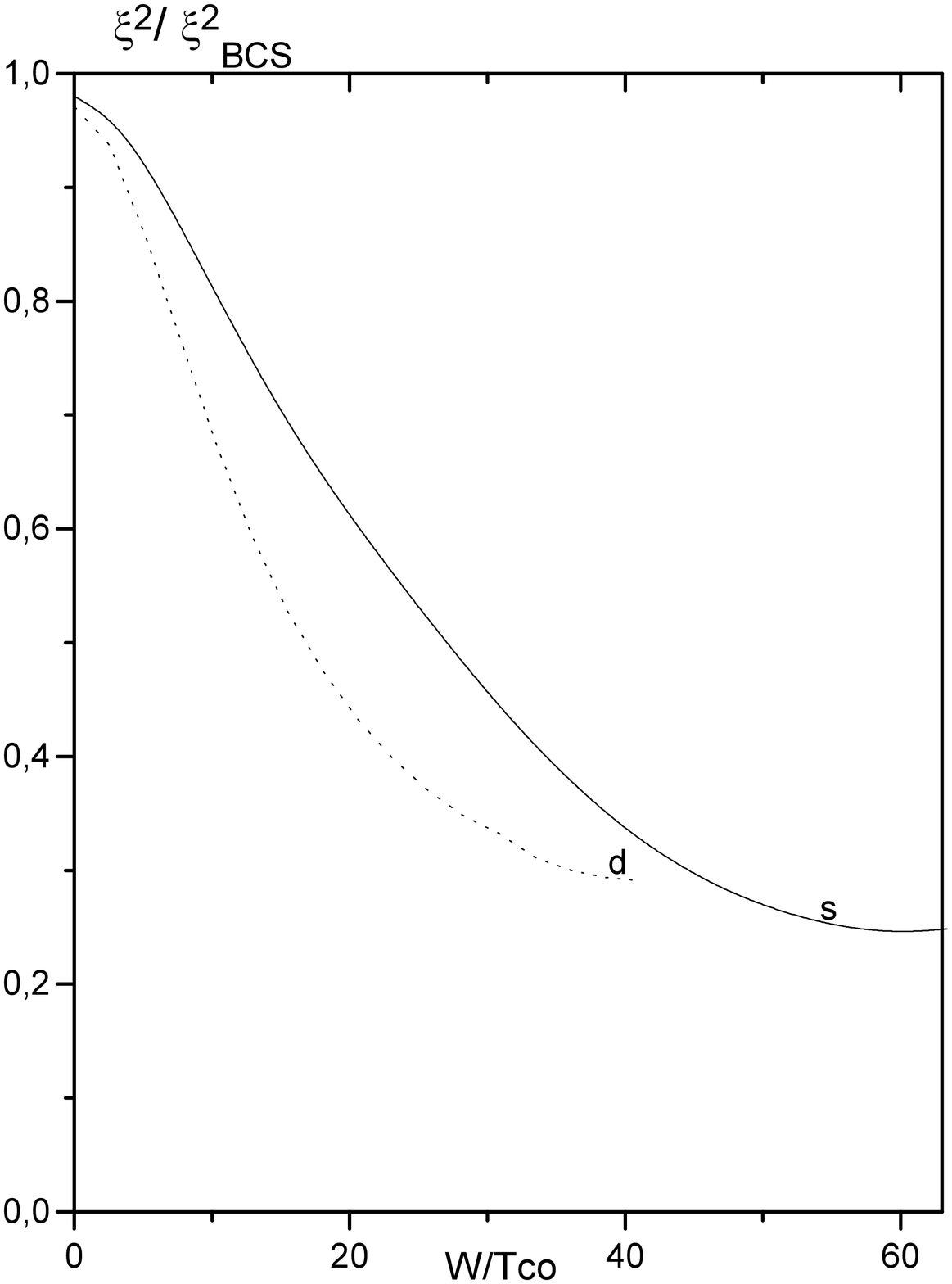}
\caption{Dependence of the square of the coherence length $\xi^2/\xi^2_{BCS}$ 
on the effective width of the pseudogap $W/T_{c0}$ for $s$-wave pairing and
scattering by charge (CDW) fluctuations (full curve) and for 
$d_{x^2-y^2}$-wave pairing and scattering by spin (AFM(SDW)) fluctuations 
(dotted curve). Data are given for the value of inverse correlation length
${\kappa a}=0.2$.} 
\label{xi+} 
\end{figure}

\begin{figure}
\epsfxsize=14cm
\epsfysize=16cm
\epsfbox{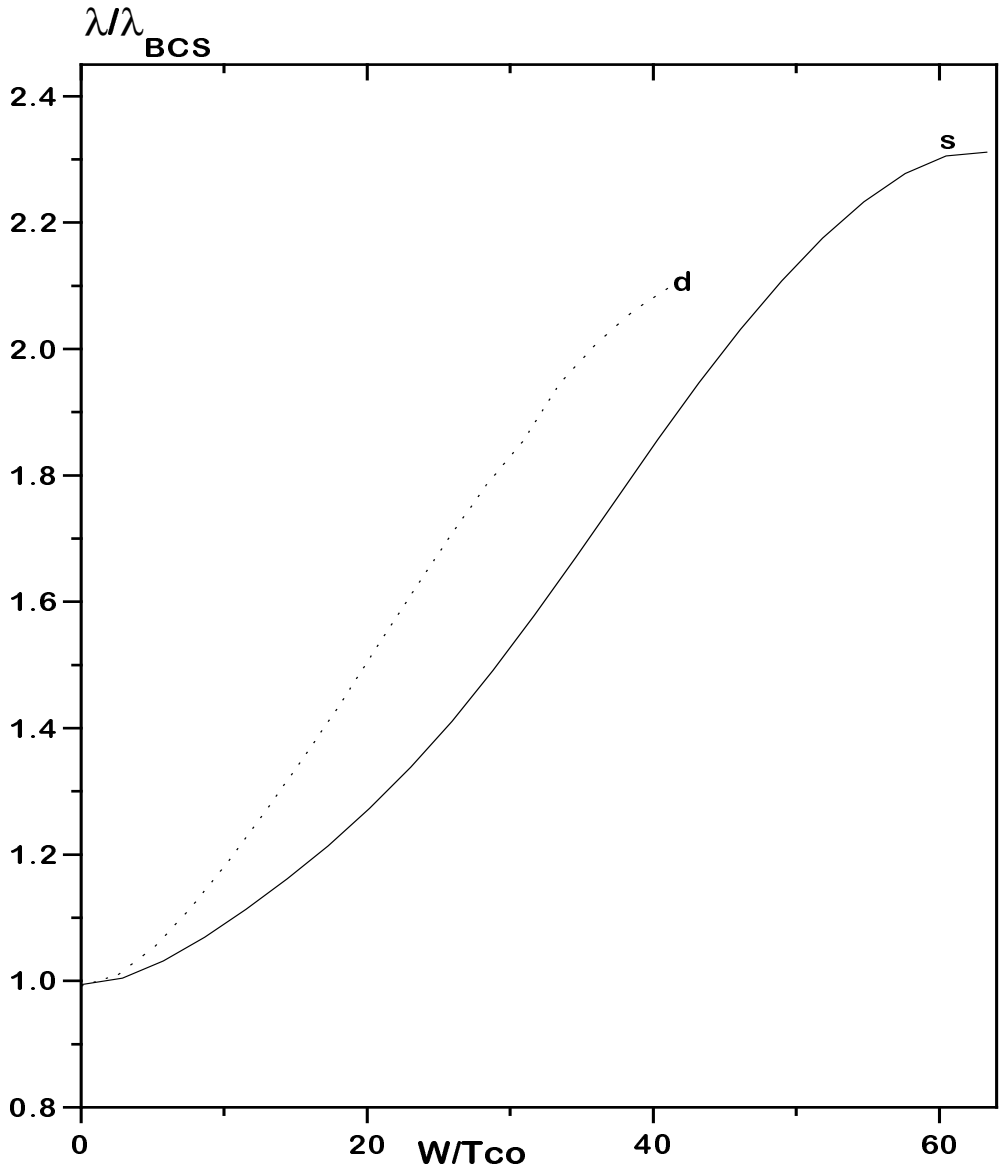}
\caption{Dependence of penetration depth $\lambda/\lambda_{BCS}$ on the
effective width of the pseudogap $W/T_{c0}$ for $s$-wave pairing and
scattering by charge (CDW) fluctuations (full curve) and for 
$d_{x^2-y^2}$-wave pairing and scattering by spin (AFM(SDW)) fluctuations 
(dashed curve). Data are given for the value of inverse correlation length
${\kappa a}=0.2$.} 
\label{lamb+} 
\end{figure}

\begin{figure}
\epsfxsize=14cm
\epsfysize=16cm
\epsfbox{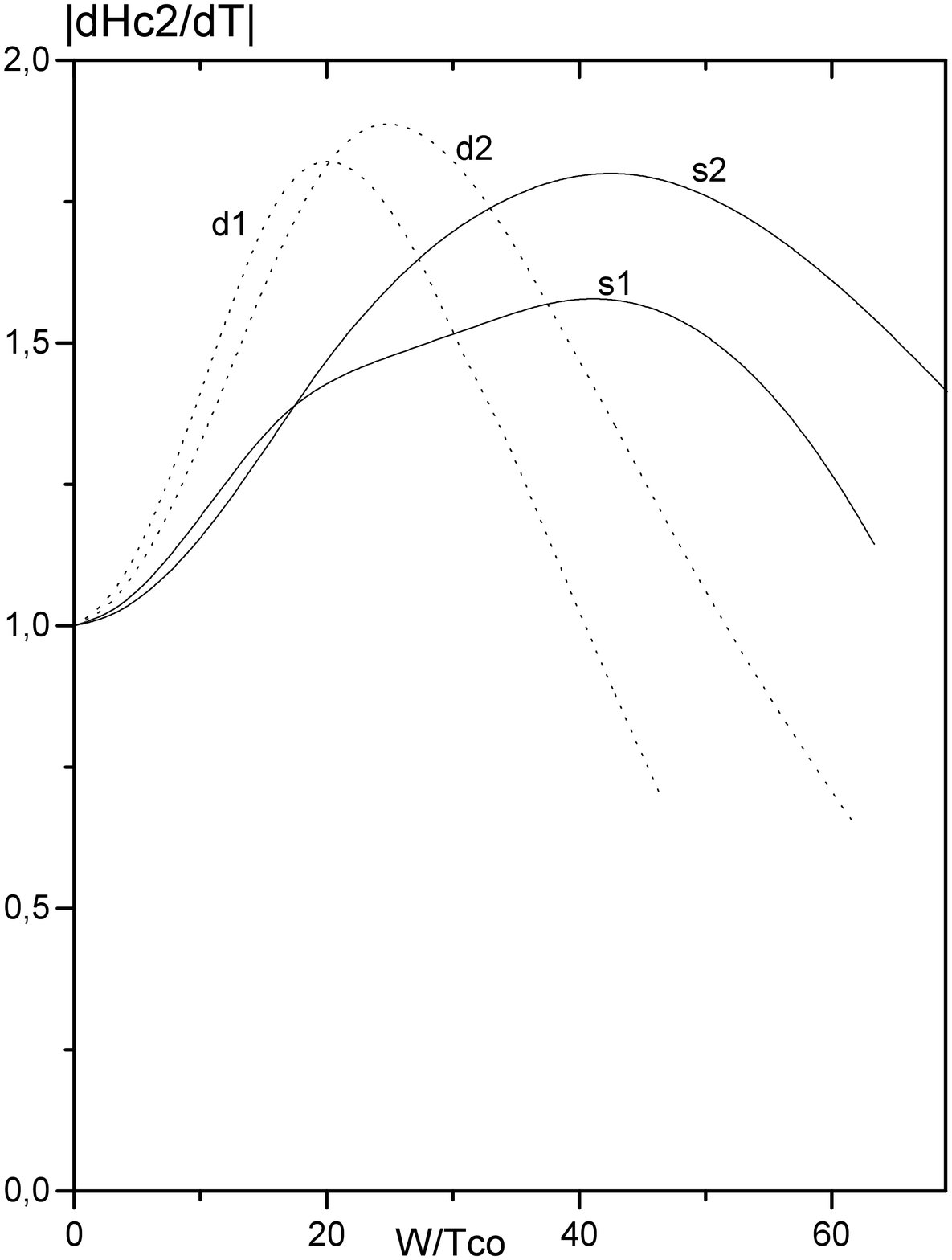}
\caption{Dependence of the derivative (slope) of the upper critical field
(normalized by its value for $W=0$) on the effective width of the pseudogap
for $s$-wave pairing and scattering by charge (CDW) fluctuations (curves s1 
and s2) and for $d_{x^2-y^2}$-wave pairing and scattering on spin (AFM(SDW)) 
fluctuations (curves d1 and d2). Data are given for the values of inverse
correlation length ${\kappa a}=0.2$ (s1 and d1) and ${\kappa a}=0.5$ (s2 
and d2).}  
\label{Hc2+} 
\end{figure}

\begin{figure}
\epsfxsize=14cm
\epsfysize=16cm
\epsfbox{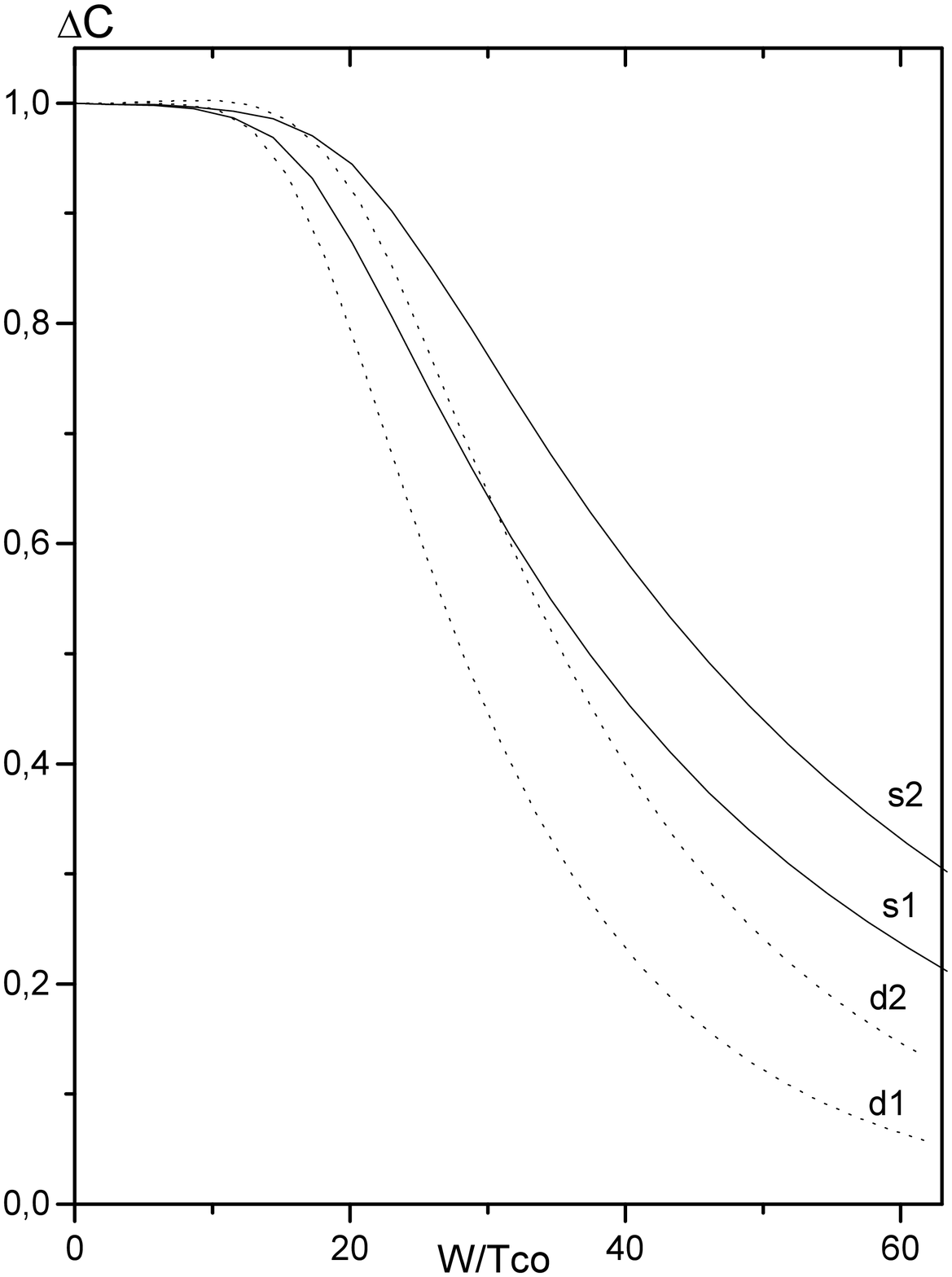}
\caption{Dependence of specific heat discontinuity at superconducting 
transition on the effective width of the pseudogap $W/T_{c0}$ for $s$-wave
pairing and scattering by charge (CDW) fluctuations (curves s1 and s2) and for
$d_{x^2-y^2}$-wave pairing and scattering by spin (AFM(SDW)) fluctuations 
(curves d1 and d2). Data are given for the values of inverse correlation length
${\kappa a}=0.2$ (s1 and d1) and ${\kappa a}=0.5$ (s2 and d2).}  
\label{DC+} 
\end{figure}

\begin{figure}
\epsfxsize=14cm
\epsfysize=16cm
\epsfbox{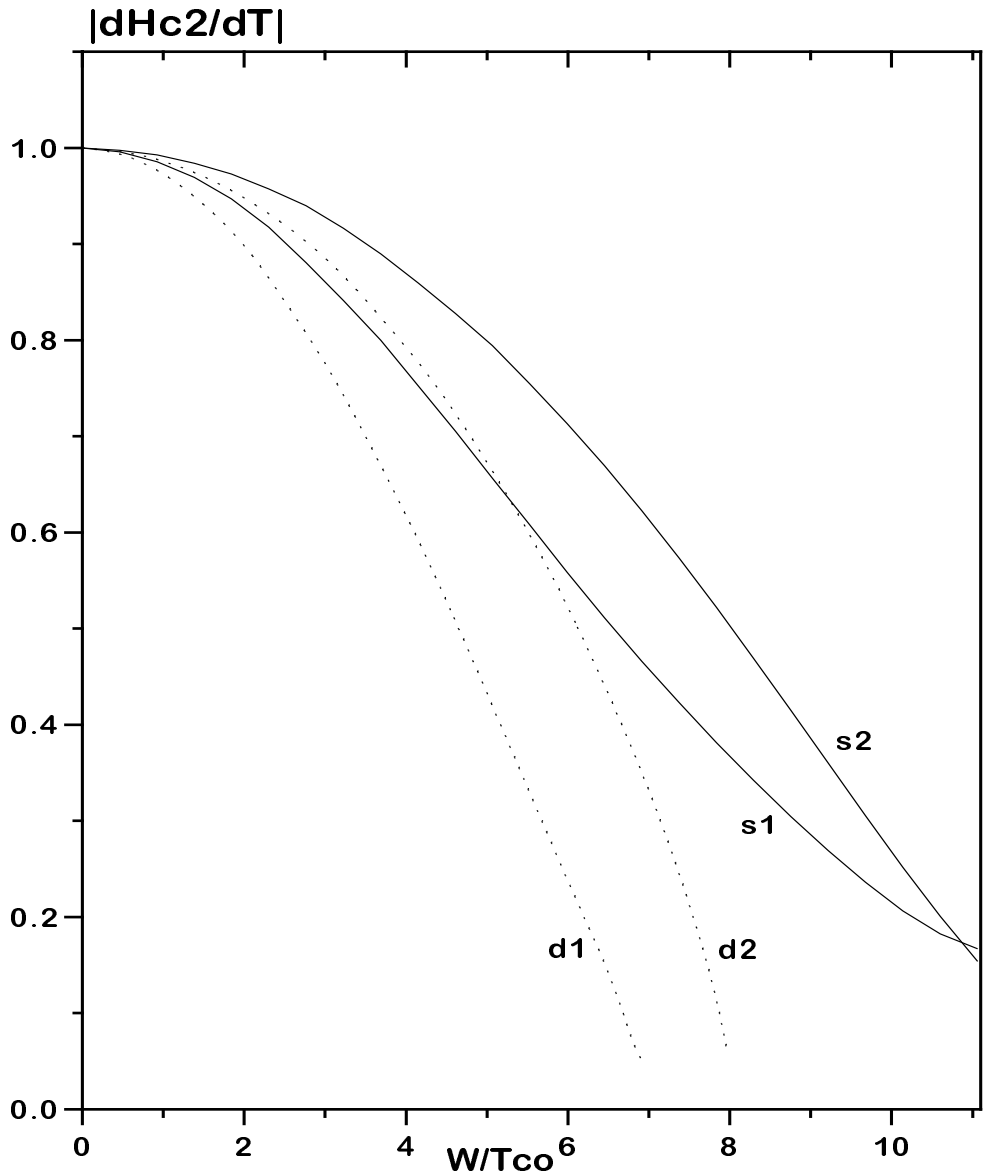}
\caption{Dependence of the derivative (slope) of the upper critical field
(normalized by its value for $W=0$) on the effective width of the pseudogap
$W/T_{c0}$ for $s$-wave pairing and scattering by spin (AFM(CDW)) fluctuations
(curves s1 and s2) and for $d_{x^2-y^2}$-wave pairing and scattering by charge
(CDW) fluctuations (curves d1 and d2). Data are given for the values of inverse
correlation length ${\kappa a}=0.2$ (s1 and d1) and ${\kappa a}=1.0$ (s2 and 
d2).}
\label{Hc2-} 
\end{figure}

\begin{figure}
\epsfxsize=14cm
\epsfysize=16cm
\epsfbox{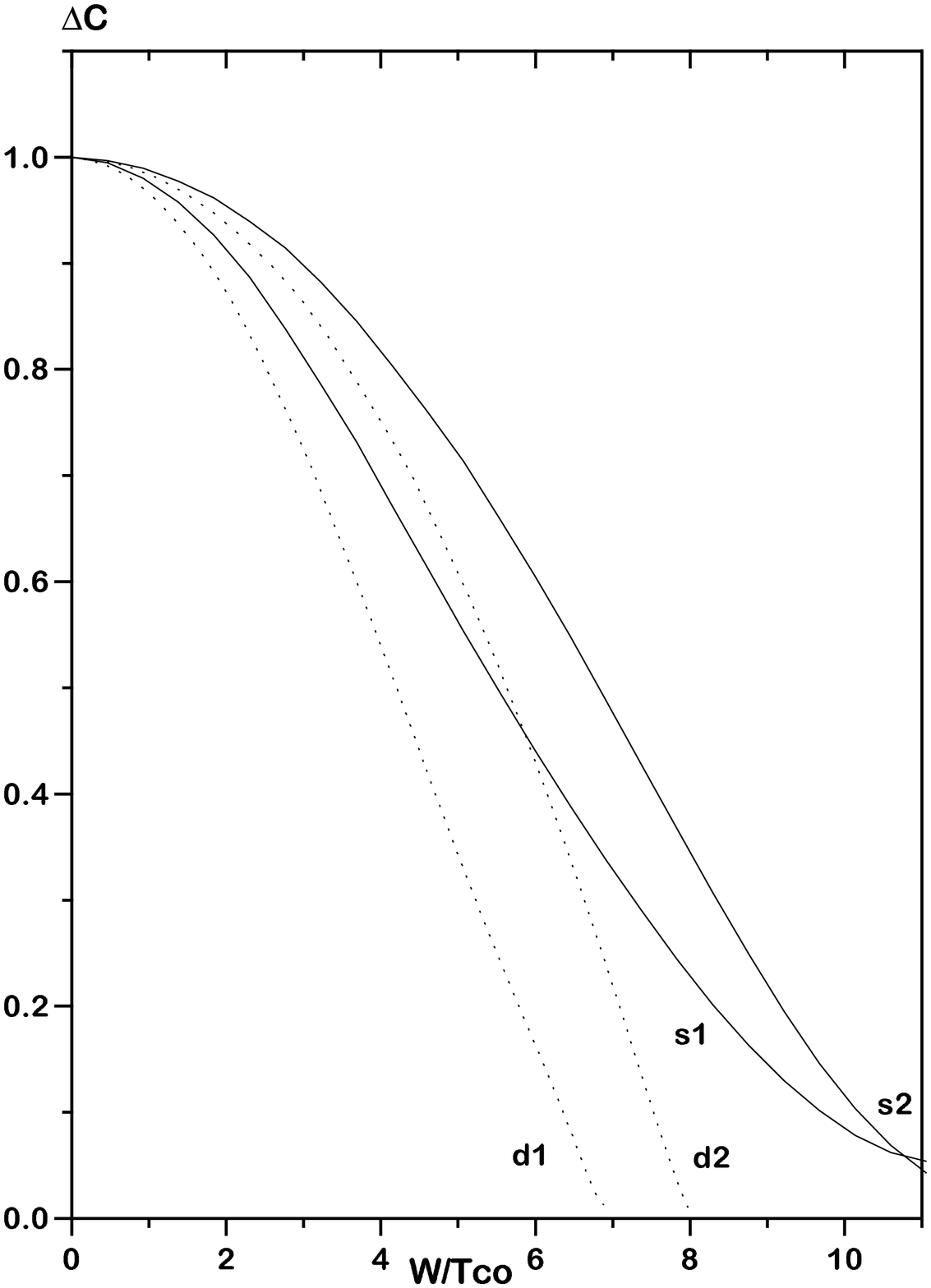}
\caption{Dependence of specific heat discontinuity at superconducting 
transition on the effective width of the pseudogap $W/T_{c0}$ for $s$-wave
pairing and scattering by spin (AFM(CDW)) fluctuations (curves s1 and s2) and 
for $d_{x^2-y^2}$-wave pairing and scattering on charge (CDW) fluctuations
(curves d1 and d2). Data are given for the values of inverse correlation length
${\kappa a}=0.2$ (s1 and d1) and ${\kappa a}=1.0$ (s2 and d2).}  
\label{DC-} 
\end{figure}

Coefficient $B$ defined by the diagram shown in Fig. \ref{GL}(b), in the
absence of pseudogap fluctuations ($W=0$) and for ${\bf q}=0$ takes the
following form:
\begin{eqnarray}
B_0=T_c\sum_{\varepsilon_n}\sum_{\bf p}
\frac{1}{(\varepsilon_n^2+\xi_{\bf p}^2)^2}e^4({\bf p})=
T_c\sum_{\varepsilon_n}\int_{-\infty}^{\infty}d\xi
\frac{1}{(\varepsilon_n^2+\xi^2)^2}
\sum_{\bf p}\delta(\xi-\xi_{\bf p})e^4({\bf p})\approx
\nonumber\\
\approx N_0(0)T_c\sum_{\varepsilon_n}\int_{-\infty}^{\infty}d\xi
\frac{1}{(\varepsilon_n^2+\xi^2)^2}
\frac{\sum_{\bf p}\delta(\xi_{\bf p})e^4({\bf p})}{N_0(0)}=
B_{BCS}<e^4({\bf p})>
\label{Bo}
\end{eqnarray}
where 
\begin{equation}
B_{BCS}=N_0(0)\frac{7\zeta(3)}{8\pi^{2}T_c^2} 
\label{Bbcs}
\end{equation}
is the standard expression for $B$ in case of isotropic $s$-wave pairing.

\newpage

\newpage

\newpage

\newpage

\newpage

\newpage

\newpage

\newpage

\newpage

\newpage

\newpage

\newpage

\newpage

\end{document}